# Linac activation of radioisotopes and underground gammaspectrometric analyses

Patrick WEBER[1*], Jean-Luc VUILLEUMIER[2], Geoffroy GUIBERT[1], Claire TAMBURELLA[1]
[1]Hôpital Neuchâtelois, Service de Radio-oncologie, Rue du Chasseral 20, CH-2300 La Chaux-de-Fonds
[2]Albert Einstein Center for Fundamental Physics, LHEP, University of Bern, Sidlerstr. 5, CH-3005 Bern, Switzerland
*Corresponding author. E-mail: Patrick.Weber2@h-ne.ch

**Abstract** - After irradiating various medical linac parts with photon beams, we performed underground gamma-ray spectroscopy (GRS) measurements of the samples only a few minutes after the irradiations, in order to observe activated short-lived radioisotopes.

Three samples were used: an Elekta Synergy flattening filter never irradiated before, an old flattening filter, unmounted from a linac in 2012 and a tungsten leaf, unmounted from the multileaf collimator (MLC) of the same linac in 2012. All three samples were measured before the irradiations in the underground laboratory of la Vue-des-Alpes.

A residual activity of 236.9 Bq was measured in the old filter with long-lived radioisotopes; with this work, it was possible to identify them: $^{57}$Co ($T_{1/2}$=271.8 days), $^{54}$Mn ($T_{1/2}$=312.3 days) and $^{60}$Co ($T_{1/2}$=5.27 years). Prior to any irradiation, the tungsten leaf and the new filter were gamma counted, and showed only tiny amounts of trace impurities of $^{238}$U, $^{57}$Co, $^{54}$Mn and $^{60}$Co, with a total activity of 0.51 and 0.02 Bq, respectively.

The new flattening filter and the leaf were then irradiated with ~200 Gy, first under 15MV and then with 6MV photon beams. The gamma counting began 20 minutes after the irradiations.

Irradiations with 15MV led to (n,γ) activation of short-lived isotopes : In the leaf, the measured activity was 1556 Bq just after the irradiation and we observed various gamma lines from $^{187}$W, $^{57}$Ni and $^{56}$Mn. In the new filter, the activity was 1097 Bq and the gamma signature of $^{56}$Mn, $^{56}$Ni, $^{57}$Ni and $^{59}$Co was clearly present. The germanium detector allowed to measure gamma lines with relative intensities (branching ratio BR) down to ~0.1%, thank to its very low background.

Irradiations of the new flattening filter at 6MV led to a small activation of $^{56}$Mn.

Gammaspectroscopic data was taken several times after the irradiation to monitor the time evolution and the total activity.

Treating patients with 15MV photon beams activate long-lived radioisotopes in the linac head. With this work, it was possible to identify these isotopes, and especially highlight the created short-lived radioisotopes.

In the literature, portable germanium detectors have been placed under the linac heads, but they were measuring the whole gamma spectra, and unable to decorrelate the activation of single linac components, or the activation of recent irradiations.

# INTRODUCTION

The radio-oncology patient treatments are realized mainly with photon beams (less frequently with electrons) with energies comprised between 4MV and 15MV in our center, and with high dose rates up to 20 Gy/min. The components of the linacs head are thus exposed for years to photon or electron beams. Activation of the chemical elements of these materials leads to unstable isotopes, decaying with very different half-lives. Most of the activated radioisotopes are decaying with short half-lives, in the range of hours or days, but some other radioisotopes have half-lives of several years. The governmental offices for radioprotection always require a measurement of the activity of the linacs, when hospitals are dismantling and/or replacing their equipments. This was the starting point of this work, by the way: the components of an old linac were measured to be still radioactive 2 years after disabling a linac and thus stored with radioactive sources. But the measurement of the activity with a portable surface activimeter is not obvious, since the residual activity is relatively low and just above the detector's background.

The purpose of this work was to measure this residual activity with an increased sensitivity on one hand, but also to identify the remaining radioisotopes present in the stored linac components.

Moreover, we also exposed to photon beams a brand new linac part (a flattening filter) with only a tiny residual radioactivity, in order to activate radioisotopes.

# MATERIALS AND METHODS

**Samples**

Three samples have been used in this work (see Figure 1). They all come from the head of Elekta linacs. The first ones were flattening filters. A flattening filter is placed in the photon beam in order to flatten the dose distribution all over the field surface. The third one was a tungsten leaf. A leaf is part of a multi-leaves collimator (MLC), used to shape the beam around the target volume, while the linac gantry is turning around the patient. The samples are listed below :

1. An old flattening filter made of stainless steel BS970-304S31. The chemical composition of the main elements is the following : Fe 67.8%, Cr 19.0%, Ni 8.0%, Mn 2.0%, S 2.0%, Si 1.0%, C 0.12%, P 0.05%. This filter was unmounted from an Elekta Precise linac in June, 2012. It has been irradiated everyday with photons up to 15MV during 10 years, between 2002 and 2012.
2. A new flattening filter from an Elekta Synergy linac head. This filter is made of the same stainless steel (BS970-304S31), same dimensions, but was never irradiated before.
3. A tungsten leaf, from the multileaf collimator (MLC), unmounted from an Elekta Precise linac in June, 2012. The chemical composition is the following : W 95.0%, Ni 2.5%, Fe 2.5%.

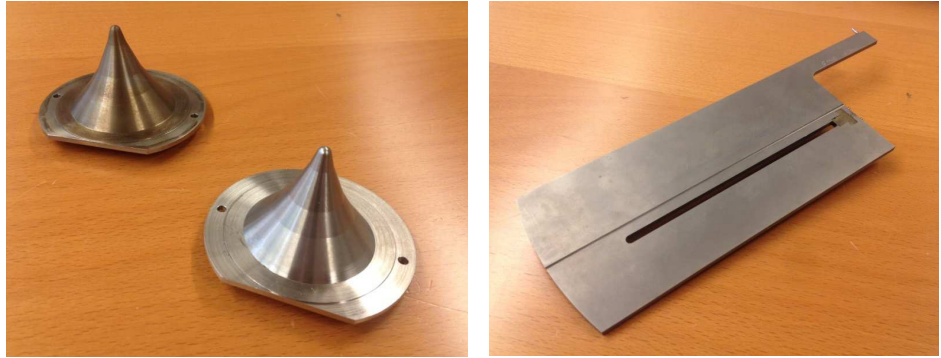

*Figure 1:* Left: old and new flattening filters. Right: Leaf from the MLC

All the samples were measured before the irradiations in the underground laboratory of la Vue-des-Alpes. Their overall residual activity has been measured by GRS. The total counts of the gamma spectra were divided by the corresponding measurement times:

1. 236.9Bq for the old filter (1'925'620 counts in 8'129 sec)
2. 0.0156Bq for the new filter (1'370 counts in 87'776 sec)
3. 0.51Bq for the leaf (8'220 counts in 16'208 sec)

**Medical linear accelerators (Linac)**
Most of the time, the patients are treated in radiation oncology with MeV photon beams. These photons are generated by bremstrahlung of high energy electrons, emitted in an electron gun and accelerated in a wave guide. A typical photon energy distribution of a 6MV beam after the flattening filter is illustrated in the Figure 2 (from Elekta Monaco Dose Calculation Technical Reference). For our Elekta Agility linacs, the nominal acceleration potentials (NAP), measured according to the AAPM TG21 are the following:

| Photon beams | NAP [MV] |
|---|---|
| 4X | 4.5 |
| 6X | 6.3 |
| 6FFF | 5.9 |
| 15X | 13.7 |

Actually, the reference dosimetry of the linacs is not based on the energy distribution, but rather on parameters extracted from the percentage depth dose (PDD) and the beam profiles, measured in particular reference conditions, detailed in national recommendations(http://ssrpm.ch/publications-and communication/recommendations-and-reports/). A modeling of the various electron and photon beams in the treatment planning system (TPS) - according to the measured profiles - allow the calculation of the dose propagation in the patient.

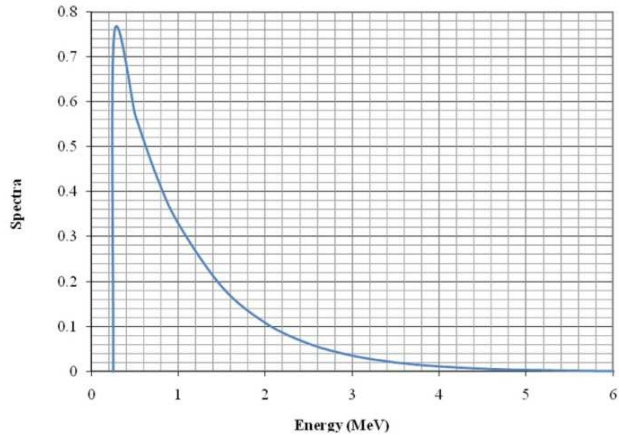

*Figure 2 :* *medical linac typical photon spectrum.*

The Figure 3 shows how the photon beam is shaped in the linac head. To every beam corresponds a single combination of difference and secondary filters. A carrousel allows the machine to select the right filters. The 6FFF is a 6MV free flattening filter photon beam, i.e without any filter. The 6FFF beam has thus a sharp profile, but with a much higher doserate, up to 20 Gy/min.
An ion chamber measures the dose and triggers the linac.
The multi-leaves collimator (MLC) has 180 leaves on the Agility head. Every leaf can move independently and shape an irradiation section that continuously moves during the treatment delivery. The MLC leaves move in the X direction, while a diaphragm moves in the Y direction.
During a VMAT treatment (Volumetric Modulated Arc Therapy), which is now the standard treatment in radio-oncology, the linac gantry turns all around the patient (1 to 4 complete or partial arcs). The MLC leaves and the diaphragm are always moving to shape the delivered dose, the doserate is varying, as well as the gantry angular speed.

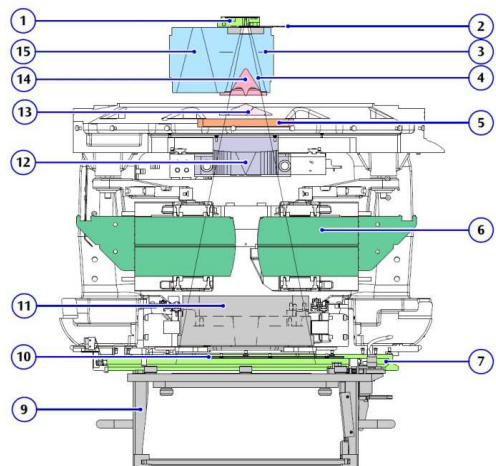

(1) Target block
(2) Primary filter assembly
(3) Primary collimator
(4) Port 1
(5) Ion chamber
(6) MLC leaves
(7) Accessory ring
(8) Electron applicator
(9) Shadow tray
(10) Mylar crosswire screen
(11) Y diaphragms
(12) Motorized wedge
(13) Secondary filter
(14) Difference filter
(15) Port 2

*Figure 3 :* *Diagram of the Elekta Agility head. Elekta Limited courtesy*

Only a few publications describe the induced radioactivity of a medical linac after the clinical beams [Fischer et al., 2006]. A germanium detector measured a gamma spectrum (see Figure 4) under a Varian TrueBeam linac [Konefal et al., 2016] several hours after therapeutic beams. 11 radioisotopes were identified: $^{56}$Mn, $^{122}$Sb, $^{124}$Sb, $^{131}$Ba, $^{82}$Br, $^{57}$Ni, $^{57}$Co, $^{51}$Cr, $^{187}$W, $^{24}$Na and $^{38}$Cl.

Nickel and manganese are some components of stainless steel, widely used in the construction of medical linacs, and particularly the flattening filter. Antimony, barium, bromine, chromium and chlorine are added to some alloys in the linacs, but should not be present in our samples.

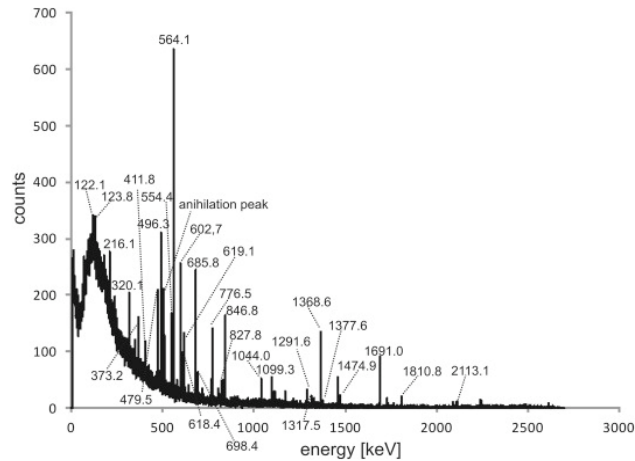

*Figure 4 : gamma spectrum measured under a Varian TrueBeam medical linac.*

**Samples irradiations**

**a)** The old filter has been irradiated for 10 years in an Elekta Precise linac, with photon beams up to 15MV. The aim was not to submit it to a new irradiation, but rather to observe long-lived radio-isotopes that remain present in such linac parts for years. The GRS analysis was thus performed in April, 2015, 22 months after its last irradiation in the Precise linac.

**b)** The new filter was irradiated with a photon beam of 15MV, a field of 15x15cm², a source-sample distance of 100cm, and a dose of 200 Gy. GRS started in the underground laboratory 17 minutes after the irradiation end (=$T_0$). Gamma spectra were acquired 27 minutes, 37 minutes, 62 minutes, 15 hours, 20 hours, 86 hours and 94 hours after $T_0$.

5 days after the 15MV irradiation, the new filter was exposed to a new irradiation of 200 Gy, but at 6MV, under the same conditions.

**c)** The leaf was irradiated with a photon beam of 15MV, a field of 15x15cm², a source-sample distance of 100cm, and a dose of 200Gy. GRS started in the underground laboratory 17 minutes after the irradiation end. Gamma spectra were acquired 23 minutes, 28 minutes, 44 minutes, 3.5 hours, 15 hours, 25 hours and 170 hours after $T_0$.

The most probable reactions produced with a high energy photon beam are photonuclear (γ,n) reactions. (γ,2n) reactions occur preferably at higher energies. The probability for (γ,p) reactions is also low, because of the proton charge. The neutron flux measured in linacs head, due to (γ,n) and (γ,2n) reactions, is described in [Konefal et al., 2012]. For an Elekta Synergy, the neutron flux at 15MV is about $1.4 \cdot 10^5$ Gy$^{-1}$. Under the high energy photon beams, radio-isotopes in our samples were activated through photonuclear (γ,n) reactions, followed by simple neutron captures (n,γ). For example, the natural Mn in the filters was activated through the reaction: n+$^{55}$Mn → $^{56}$Mn + γ
The gamma-ray spectroscopy was performed to measure these photons.

**Gamma-ray spectroscopy**
Gamma-ray measurements of radionuclides were performed with an ultra-low noise germanium detector, described previously in [Gonin et al. 2003, Leonard et al. 2008, Weber et al. 2017]. It was used mostly for material screening for the Enriched Xenon Observatory (EXO) experiment, searching for neutrinoless double beta decay in $^{136}$Xe. The detector itself, made by Eurisys-Mesures, is a 400 cm$^3$ single germanium crystal, p type coaxial. The energy resolution is 1.4 keV FWHM at 238 keV, and 2.5 keV at 1460 keV, scaling with the square root of the energy above that.

## RESULTS

**Old filter**
Two years after the disabling of the Precise linac, a residual activity of the flattening filter for 15MV of 236.9Bq was measured. The gamma spectrum of this filer is shown in Figure 5. Long-lived radio-isotopes remained in the filter and were still measured. The Table 1 summarizes the measured elements, essentially $^{57}$Co, $^{54}$Mn and $^{60}$Co.

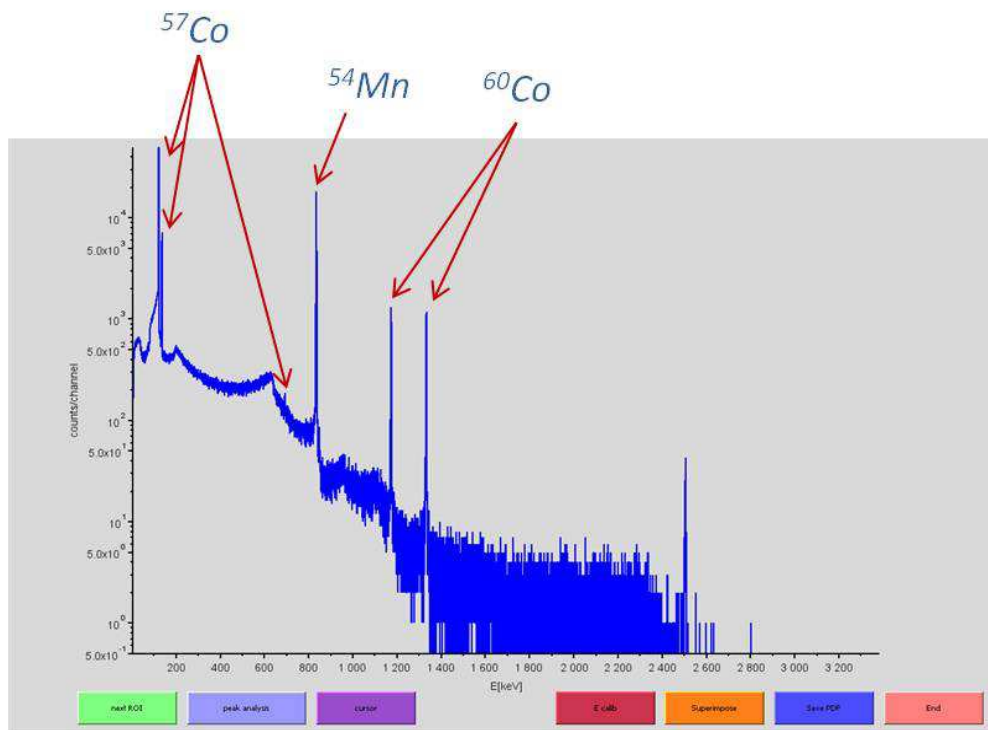

*Figure 5 : GRS spectrum of the old flattening filter.*

| Observed Gamma lines E [keV] | Element | $T_{1/2}$ | Branching ratio BR [%] |
|---|---|---|---|
| 122.1 | $^{57}$Co | 271.8 d | 85.6 |
| 136.5 | | | 10.7 |
| 692 | | | 0.16 |
| 834.8 | $^{54}$Mn | 312.3 d | 99.97 |
| 1173.2 | $^{60}$Co | 5.27 y | 99.98 |
| 1332.5 | | | 99.97 |
| 2505.7 | | | sum |

*Table 1 :* *Long-lived radioisotopes measured in the old filter.*

**New filter, 15MV**
The new flattening filter was never irradiated before and was analyzed by GRS prior to any irradiations. A total intrinsic radioactivity of 0.0156Bq has been measured. The germanium detector itself has a total background of 0.0065Bq. The Table 2 shows trace impurities naturally present in the new filter. Isotopes from the $^{238}$U decay chain and $^{40}$K were observed.

The filter was then irradiated with 200Gy at 15MV under the linac. A broad variety of short-lived isotopes was activated, as listed in Table 3. It is interesting to notice that it was possible to observe gammas emitted with branching ratios down to 0.1%, thank to the very low detector background. The total activity was decreasing exponentially with time, as shown on the Figure 6. The total activity at the end of the irradiation ($T_o$) has been estimated to 1050Bq, according to the Figure 7.

| E [keV] | Element | BR [%] |
|---|---|---|
| 295.2 | $^{238}$U($^{214}$Pb) | 18.5 |
| 351.9 | | 35.8 |
| 609.3 | $^{238}$U($^{214}$Bi) | 44.8 |
| 1120.3 | | 14.8 |
| 1764.5 | | 15.4 |
| 1460.8 | $^{40}$K | 11.5 |

*Table 2 :* *Trace impurities measured in the new filter, before any irradiation.*

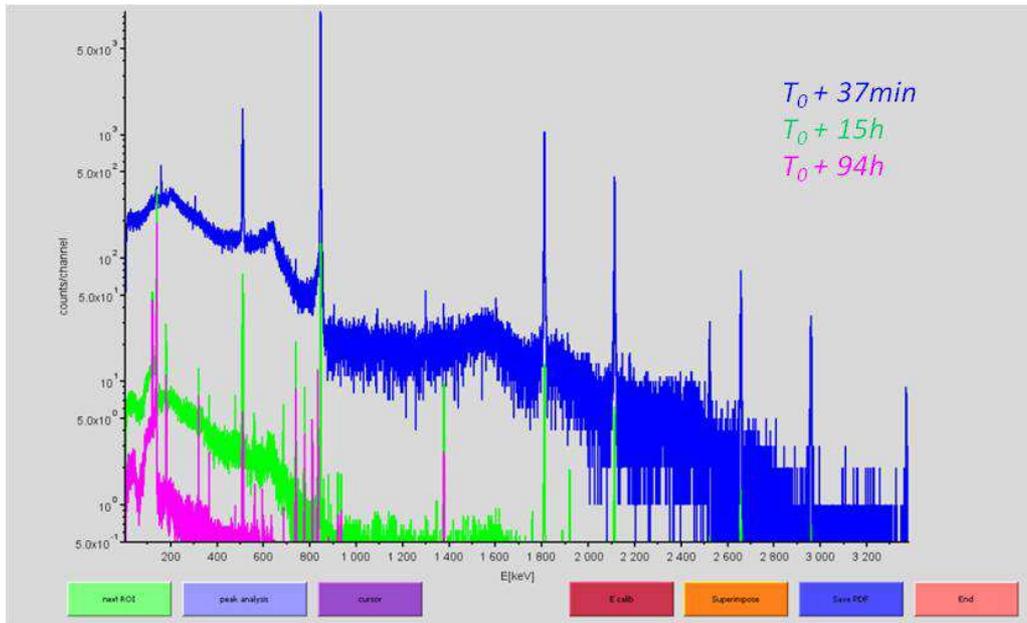

*Figure 6* : *GRS spectra of the new filter, after irradiation with 200Gy at 15MV, as a function of time after the irradiation end.*

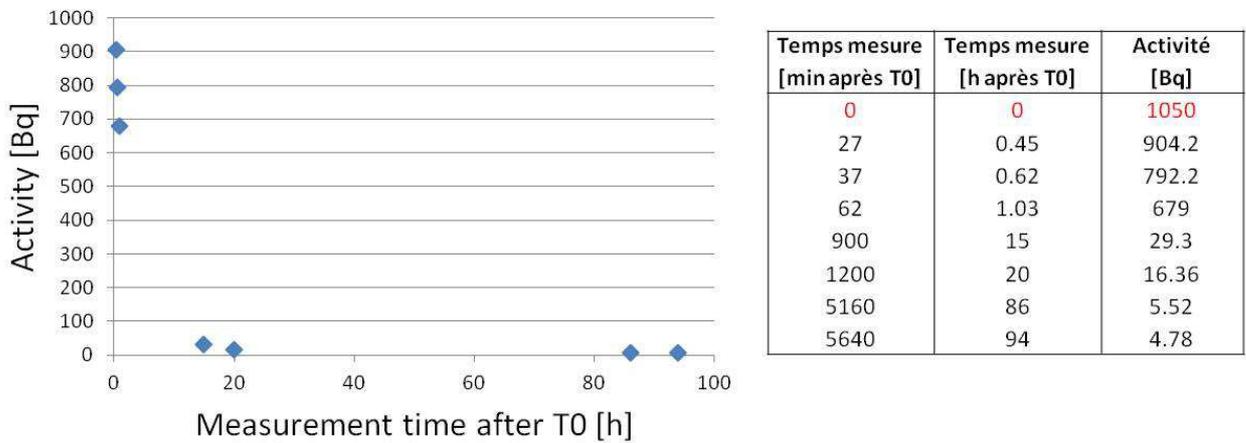

*Figure 7* : *Total activity of the new filter after irradiation with 200Gy at 15MV, as a function of time after the irradiation end.*

| Observed lines E [keV] | Element | $T_{1/2}$ | BR [%] |
|---|---|---|---|
| 122.1 | $^{57}$Co | 271.8 d | 85.6 |
| 136.5 | | | 10.7 |
| 127.2 | $^{57}$Ni | 35.6 h | 16.7 |
| 1377.6 | | | 81.7 |
| 140.5 | $^{99}$Mo | 65.9 h | 4.5 |
| 181.1 | | | 6.1 |
| 739.5 | | | 12.1 |
| 777.9 | | | 4.3 |
| 320.1 | $^{51}$Cr | 27.7 d | 10 |
| 366.3 | $^{65}$Ni | 2.52 h | 4.8 |
| 480.4 | $^{56}$Ni | 6.08 d | 36.5 |
| 811.8 | | | 86 |
| 834.8 | $^{54}$Mn | 312.3 d | 100 |
| 846.8 | $^{56}$Mn | 2.58 h | 98.9 |
| 1810.7 | | | 27.2 |
| 2113.1 | | | 14.3 |
| 2522.9 | | | 0.99 |
| 2657.5 | | | 0.65 |
| 2959.8 | | | 0.31 |
| 3369.6 | | | 0.17 |
| 1099.3 | $^{59}$Fe | 44.5 d | 56.5 |
| 1291.6 | | | 43.2 |
| 1173.2 | $^{60}$Co | 5.27 y | 99.9 |
| 1332.5 | | | 100 |
| 935.5 | $^{52}$Mn | 5.59 d | 94.9 |
| 1377.6 | $^{57}$Ni | 35.6 h | 81.7 |
| 1757.6 | | | 5.8 |
| 1919.5 | | | 12.3 |

*Table 3 : Radioisotopes measured in the new filter after irradiation with 200Gy at 15MV*

**New filter, 6MV**
The irradiation of the new filter with 200Gy at 6MV led to a small increase of the total activity, from 2.46Bq before the 6MV irradiation to 4.60Bq after. At 6MV, only elements with low energy photonuclear absorption thresholds can be activated. We measured gammas from $^{56}$Mn, at 846.8keV (see Figure 8), 1810.8keV and 2113.1keV.

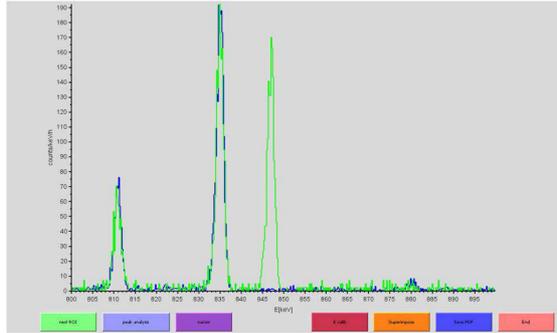

*Figure 8 :* Green: 846.8keV gamma from the $^{56}$Mn decay, after irradiation with 200Gy at 6MV. Blue: spectrum before 6MV irradiation.

**Leaf**

The total activity of the leaf was 0.51Bq before the irradiation at 15MV. According to the curve fit in the Figure 9, it has been estimated to be 1805Bq at the end of the irradiation with 200Gy ($T_o$). The total activity decreased exponentially after $T_o$. The compound radio-isotopes half-life can be calculated:

$$A(t) = A_0 \cdot e^{-(\lambda \cdot t)}, \text{ with } \lambda = \frac{\ln(2)}{T_{1/2}}$$

$$\Rightarrow T_{1/2} = \frac{\ln(2)}{\lambda} = 24.8\ h$$

A broad variety of short-lived isotopes have been activated by neutron capture during the irradiation, with a lot of gammas emitted in the $^{187}$W decay process, that have a short $T_{1/2}$=23.72 hours. As $^{187}$W decays dominates, it is logical that the compound half-life is slightly above the $^{187}$W half-life. The Table 4 summarizes the observed radio-isotopes.

Photons measured around 57.9keV and 67.6keV are probably a mixing of x-rays, originated from electrons conversions.

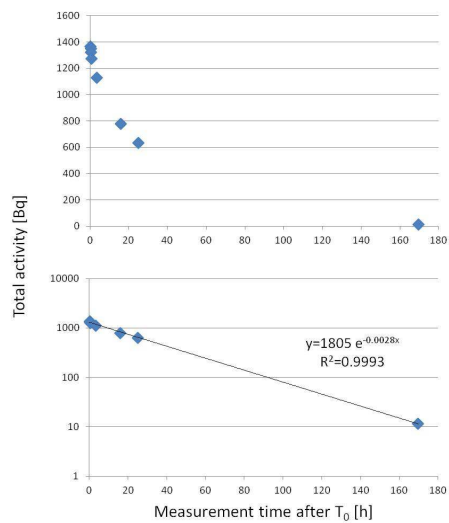

*Figure 9 :* Total activity of the tungsten leaf, after irradiation with 200Gy at 15MV, as a function of time after the irradiation end.

| E [keV] | Element | $T_{1/2}$ | BR [%] |
|---|---|---|---|
| 57.9 | KL | --- | X-ray |
| 67.6 | KM | --- | X-ray |
| 72 | | | 11.14 |
| 134.2 | | | 8.85 |
| 206.2 | | | 0.14 |
| 246.3 | | | 0.12 |
| 479.5 | | | 21.8 |
| 551.5 | | | 5.08 |
| 589.1 | | | 0.12 |
| 618.4 | $^{187}W$ | 23.72 h | 6.28 |
| 625.5 | | | 1.09 |
| 685.8 | | | 27.3 |
| 745.2 | | | 0.3 |
| 772.9 | | | 4.12 |
| 816.5 | | | 0.01 |
| 864.6 | | | 0.34 |
| 879.4 | | | 0.14 |
| 122.1 | $^{57}Co$ | 271.8 d | 85.6 |
| 136.5 | | | 10.7 |
| 810.8 | $^{58}Co$ | 70.86 d | 99 |
| 834.8 | $^{54}Mn$ | 312.3 d | 99.98 |
| 846.8 | $^{56}Mn$ | 2.58 h | 98.9 |
| 1809.6 | | | 27.2 |
| 1173.2 | $^{60}Co$ | 5.27 y | 99.9 |
| 1332.5 | | | 100 |
| 1377.6 | $^{57}Ni$ | 35.6 h | 81.7 |
| 1919.5 | | | 12.3 |
| 1480.8 | sum | | |

*Table 4 : Radioisotopes measured in the tungsten leaf after irradiation with 200Gy at 15MV*

## DISCUSSION

Treating patients with 15MV photon beams, as it is done daily in the clinical routine, activates long-lived radioisotopes in the linac head. With this work, it was possible to identify them : a residual activity of 236.9Bq was measured in the old filter almost two years after the linac dismantling, with long-lived radioisotopes such as $^{57}Co$ ($T_{1/2}$=271.8 days), $^{54}Mn$ ($T_{1/2}$=312.3 days) and $^{60}Co$ ($T_{1/2}$=5.27 years). For that reason, avoiding long stays in the treatment room makes sense for the clinical staff.

The aim of this work was to highlight the activated short-lived radioisotopes in medical linac head's components. It was possible to distinguish which radioisotopes were created in which materials, thank to the irradiation of single components of a linac's head.

Moreover, the extremely low background of the germanium detector, in an underground laboratory, allowed to measure gammas with tiny branching ratios, down to ~0.1%.

Before the irradiations, the tungsten leaf and the new filter exhibited only trace impurities of $^{238}$U, $^{57}$Co, $^{54}$Mn and $^{60}$Co, with a total activity of 0.51 and 0.02Bq, respectively. Irradiations with 15MV led to (n,γ) activation of short-lived isotopes : In the leaf, the measured activity was 1556 Bq just after the irradiation and we observed various gamma lines from $^{187}$W, $^{57}$Ni and $^{56}$Mn. In the new filter, the activity was 1097Bq and the gamma signature of $^{56}$Mn, $^{56}$Ni, $^{57}$Ni and $^{59}$Co was clearly present. Irradiation at 6MV led to a small activation of $^{56}$Mn.

For all the samples, GRS data was taken several times after the irradiation to monitor the time evolution and the total radioactivity.